\documentclass[preprint2,10pt,a4paper]{aastex}
\usepackage{amsmath}
\usepackage{graphicx}
\begin{document}
\title{Hydromagnetic waves in a superfluid neutron star with
strong vortex pinning}
\author{Maarten van Hoven and Yuri Levin}
\affil{Leiden university, Leiden Observatory and Lorentz Institute,
 P. O. Box 9513, NL-2300 RA Leiden}
\email{vhoven@strw.leidenuniv.nl, yuri@strw.leidenuniv.nl}
\begin{abstract} \noindent
Neutron-star cores may be hosts of a unique mixture of a neutron superfluid
and a proton superconductor. Compelling theoretical arguments have been
presented over the years that if the proton superconductor is of type II,
than the superconductor fluxtubes and superfluid vortices should
be strongly coupled and
hence the vortices should be  pinned to the proton-electron
plasma in the core. We explore the effect of this pinning on the
hydromagnetic waves in the core, and discuss 2 astrophysical applications
of our results: 1. We show that even in the case of strong pinning,
the core Alfven waves thought to be responsible for the low-frequency
magnetar quasi-periodic oscillations (QPO) are not significantly mass-loaded
by the neutrons. The decoupling of $\sim 0.95$ of the core mass from
the Alfven waves is in fact required in order to explain the QPO frequencies,
for simple magnetic geometries and for magnetic fields not greater than
$10^{15}$ Gauss. 2. We show that in the case of  strong vortex pinning,
 hydromagnetic stresses exert stabilizing
influence on
the Glaberson instability, which has recently been proposed as
a potential source of superfluid turbulence in neutron stars.
\end{abstract}

\section{Introduction}
Since the late 1950's,  it has been realized that neutron-star interior
may consist of a number of quantum fluids (see Shapiro \&
Teukolsky 1983 for a review). Currently, it is thought that both neutron
superfluid and proton superconductor are likely to coexist in the
neutron-star cores (see, e.g., Link 2007 for a discussion). Several
researchers have
argued that
if the proton superconductivity were of the type II, then the superconductor
fluxtubes would couple strongly to the neutron superfluid vortices. This
line of reasoning is based on the fact  that nuclear forces contain
velocity-dependent terms, which results in the entrainment of protons
in the neutron supercurrent (Alpar, Langer,  \& Sauls, 1984). Therefore,
the vortices
are sheathed by charged currents entrained in the
superfluid flow, and are strongly magnetized. Magnetic fluxtubes  interact
strongly
with the magnetized
vortices, similar to the way in which the fluxtubes interact between
each other (Ruderman, Zhu, \& Chen 1998, and references therein).
As a result of this coupling, the vortices get strongly pinned to
the proton-electron plasma in the core.
Such pinning would have important implications for the neutron-star
phenomenology. Ruderman, Zhu, \& Chen (1998) have argued that the
vortex-pinning
in the core may be responsible for the observed glitches in the pulsar
rotation rates. Link (2003) has considered the effect
of the vortex-fluxtube interaction on the dynamics of the precessing
pulsar PSR 1828-11 (observed by Stairs, Lyne, \& Shemar 2000).
Building on the theoretical work by Shaham (1977) and Sedrakian,
Wasserman, \& Cordes
(1999), he has concluded that the interaction, if present,
would ultimately lead to the fast precession. Since
PSR 1828-11 is precessing slowly and persistently, Link (2003) has argued
that
the core vortex pinning is excluded by
the observations and hence that either the proton superconductor might be of type I,
or that both proton and neutron condensates do not coexist 
inside that pulsar. While
Link's argument is suggestive, we believe it is premature
to rule out strong vortex pinning in the cores of all neutron stars.

In this paper we consider hydromagnetic waves in the case when the neutron
vortices are
strongly pinned to the proton-electron plasma in the core. We have 2 main
astrophysical
motivations for studying this problem.

The first one is due to the fairly recent observations
of the quasi-periodic oscillations (QPOs) of the x-ray luminosity  in the
tails of
giant magnetar flares (Israel et al.~2005, Strohmayer \& Watts 2005, 2006,
see also
earlier but
lower signal-to-noise measurements of Barat et al.~1983).
Israel et al.~2005 has argued that the lowest-frequency and the
longest-lived QPO of
$\sim 18$Hz is likely
to represent an Alfven-type oscillation in the magnetar core (this
frequency is too
low to be associated
with the torsional modes of the crust). Levin (2006) has shown that for a
magnetar-strength field the
crustal motion [which is thought to be either powering power the flare
(Thompson \&
Duncan 1995) or responding
to a global reconnection event in the magnetosphere (Lyutikov 2003)] would
excite
the core Alfven waves
on the timescale of several oscillation periods. Since then, a significant
body of
theoretical work
has been devoted to a study of global magnetar vibrations, which would
involve both
hydromagnetic waves in the core
and elasto-magnetic shear waves in the crust [Glampedakis, Samuelsson, \&
Andersson
2006, Levin 2007 (from here on
L07), Sotani, Kokkotas, \& Stergioulas 2007, Lee 2007]. In particular, L07
has
argued that the long-lived
low-frequency QPOs are associated with the special spectral points of the
Afven
continuum in the magnetar core.
For simple B-field configurations these special points can be found
analytically,
and do not depend on the details of
the crust. For example, for a uniform internal B-field, the lowest QPO is
expected
at the frequency
\begin{equation}
\nu_{\rm alfven}\sim {B_{\rm eff}\over 2R\sqrt{4\pi\rho_c}}
\label{nua}
\end{equation}
where $R$ is the radius of the fluid part of the star, $\rho_c$ is the
density of the
the part of the fluid which is coupled to the Alfven waves, and $B_{\rm
eff}$ is the
effective
magnetic field which is given by\footnote{The occurence of $B_{\rm eff}$ in Eq.~(\ref{nua}) and (\ref{beff}) can be understood as follows: the magnetic tension force acting on surface $\Sigma$ perpendicular to the fluxtubes containing $N$ fluxtubes is given by $N\Delta\Sigma \cdotp B_{\rm cr}^2/4\pi = \Sigma \cdotp T_{\rm eff}$, where $\Delta\Sigma$ is the cross section of a single fluxtube, and $T_{\rm eff}$ is the effective tensile stress. The magnetic flux through $\Sigma$ is  $\Phi = \Sigma \cdotp B = N\Delta\Sigma \cdotp B_{\rm cr}$, and we find $T_{\rm eff} = B B_{\rm cr}/4\pi = B_{\rm eff}^2/4\pi$ (as opposed to $T = B^2/4\pi$, where $T$ is the corresponding part of the Maxwell stress tensor). (A detailed derivation is given in Easson \& Pethick, (1977)).}
\begin{equation}
B_{\rm eff}=\sqrt{B B_{\rm cr}}.
\label{beff}
\end{equation}
Here $B$ is the average magnetic field, $B_{\rm cr}\simeq 10^{15}$G is the
value of
the  critical
magnetic field confined to the fluxtubes. 
>From Eq.~(\ref{beff}) we see
that the
magnetar QPO frequencies
could provide an interesting constraint on the magnetic-field
strength and geometry.
However,  interaction between neutron and proton superfluids
could affect core Alfven waves, by effectively mass-loading them with
neutrons.
We shall consider the extreme case of such interaction---the strong vortex
pinning
on the fluxtubes,
and show that it does not significantly alter the Alfven-wave propagation  in
slowly-spinning magnetars
(but is important for the Alfven waves in the fast-spinning radio-pulsars).
This simplifies the interpretation of the QPO frequencies, and shows that
it is
valid to assume that
$\rho_c$ is just the density of protons, about 5\% of the total fluid
density.

Our second motivation is the
recent theoretical discussion of the superfluid turbulence in the
neutron-star cores.
Superfluid turbulence has been discussed in the context of the laboratory
Helium fluid
for the last 30 years
(see, e.g., Donnelly 1991 for a review).
In a  ground-breaking series Peralta, Melatos,
Giacobello, \& Ooi at the University of Melbourne (2005, 2006; hereafter
PMGO5 and
PMGO6)
have applied the superfluid-helium ideas to neutron-star interiors.
PMGO have developed from scratch a code which studies numerically the
2-component
superfluid
dynamics. The 2 components in PMGO are the neutron superfluid and the
normal neutron
fluid which
are coupled via the mutual friction force at the superfluid vortices;
this mixture is expected if the core temperature is a significant fraction
of the
critical
temperature of the superfluid. The equations of motions used by PMGO were
derived by
Hall and Vinen (1956)
and Bekharevich and Khalatnikov (1961). PMGO5 have studied, for the first
time, the
superfluid
spherical Taylor-Couette flow, and find that it becomes turbulent in 2
steps: 1. the
normal component
develops meridional circulation due to the Eckman pumping (see, e.g.,
Pedlosky 1987),
and 2. the component of the meridional flow of the normal fluid which is
directed
{\it along} the superfluid vortices drives the vortex Kelvin waves
unstable; this is known as  the Glaberson (or sometimes Donnelly-Glaberson)
 instability (Glaberson, Johnson, Ostermeier 1974, Donnelly 1991).
In PMGO5 simulations, the Glaberson instability leads to turbulence.
Interestingly,
PMGO6 and Melatos and Peralta (2007) demonstrate that the superfluid
turbulence could
affect the pulsar spin and could be behind the well-known pulsar timing
noise and
spin glitches.

More recently, Sidery, Andersson, and Gomer (2007, hereafter SAG), and
Glampedakis, Andersson, and Jones (2007, hereafter
GAJ1, and 2008, hereafter GAJ2)
developed an analytical theory of the Glaberson instability in neutron
stars. Their
2-component fluid
consists of the neutron superfluid and the proton superconductor, which
are, like in
PMGO,
 coupled via the mutual-friction force.
SAG have considered the limit of the weak mutual friction (see below) and
infinitely
heavy proton
superfluid, and
 found that it was the inertial waves in the neutron superfluid which were
subject
to the
Glaberson instability. GAJ1 and GAJ2 have extended
this analysis to include the regime of realistic proton-to-neutron
mass ratio and of arbitrarily strong mutual friction. Notably, they found
that
the Glaberson instability operated even in the regime of strong pinning.
But where
would
the initial relative flow between the protons and neutrons come from?  GAJ
have argued that if the pinned neutron vortices were misaligned with the
angular
velocity of
the protons, then this would naturally lead to the relative proton-neutron
flow
which would potentially be strong enough to drive the
turbulence in some parts of the star. Without turbulence, the star with
misaligned
pinned vortices would undergo fast precession (Shaham 1977) which, although probably hard to detect, has yet not been observed
in any of
the
currently-timed radio pulsars or magnetars. GAJ2 argue that this the
Glaberson-instability-induced turbulence may generically prevent the star
getting into
a configuration with the fast precession.
However, this conclusion is premature.
 One important piece of physics which is not considered in GAJ is the strong
hydromagnetic stress
inside the proton superfluid, which, as we show below, has a
suppressing
effect on the Glaberson
instability and hence on the development of the superfluid turbulence.
We will derive, however, a robust upper limit on the
angle of fast precession, which is determined by the maximum possible
mutual torque between the neutron superfluid and the
proton superconductor. The maximal precession angle turns out to be much smaller
than 1 degree, and thus it is not surprising that the fast precession has never been
observed in real neutron stars.

The plan of the paper is as follows. In the next section we derive the
dispersion
relation for
the hydromagnetic waves when the superfluid vortices are rigidly attached
to the
core plasma.
In sections 3 and 4, we consider applications to oscillating magnetars and
precessing pulsars,
respectively. We end with the general discussion in section 5.

\section{Dispersion relations}
As a starting point, we utilize the plane-wave analysis outlined in GAJ.
We follow closely the notation of and reasoning behind GAJ1's basic equations
(1)---(7), and our
derived dispersion relation is identical to their Eq.~(10) in the limit of
zero
hydromagnetic stress, but
has important extra terms when the hydromagnetic stress is included.
We begin with the two-fluid dynamical equations, cf.~ Eqs.~(1) and (2) in
GAJ1:
\begin{eqnarray}
D^{n}_{t} \vec{v}_n+\nabla \psi_n&=&2 \vec{v}_n\times
\vec{\Omega}+\vec{f}_{\rm
mf}\label{n1}\\
D^p_t \vec{v}_p+\nabla \psi_p&=&2 \vec{v}_p \times \vec{\Omega}-\vec{f}_{\rm
mf}/x_p+\nu_{\rm ee} \nabla^2
                                  \vec{v}_p\nonumber\\
                            & &+\vec{f}_{\rm hm}
\label{np}
\end{eqnarray}
Here $\vec{v}_n$ and $\vec{v}_p$ are the velocities of the neutron and proton
condensates, respectively,
the full time derivatives $D_t$ are defined in the usual way  as
$D^{n,p}_t=\partial/\partial t +\vec{v}_{n,p}\cdot \nabla$, $\psi_{n,p}$
is the sum
of specific chemical
and gravitational potentials, $\vec{f}_{\rm mf}$ is the acceleration of
the neutron
superfluid   due to the mutual friction between its vortices
and the charged plasma, $x_p=\rho_p/\rho_n$ is the density ratio
between the proton charged and neutral components of the
interior ($\sim 5$\%), $\Omega$ is the angular velocity of the rotating frame
in which all of the velocities are defined, $\nu_{\rm ee}$ is the kinetic
viscosity of the plasma due to electron-electron scattering,
 and 
\begin{equation}
\vec{f}_{\rm hm} = \vec{B}_{\rm eff} \cdotp \vec{\nabla} \vec{B}_{\rm eff} / 4\pi \rho_p
\label{induction}
\end{equation}
 is the acceleration of the plasma due to the
hydromagnetic stress. We note that because in a type-II superconductor the distance between the fluxtubes is much larger than the 
fluxtube diameter, we ignore magnetic pressure. In writing down Eq.~(\ref{n1}), we have followed GAJ
and
neglected explicitly
the effect of superfluid entrainment (which we expect will not
qualitatively change
our results) and
the individual tension force for the vortices (which is can be neglected
if the
wavelength of the waves is
much greater than the inter-vortex distance).
We shall use the following conventional form (Hall \& Vinen 1956,
Bekharevich \&
Khalatnikov 1961,
PMGO, SAG, and GAJ) for the mutual-friction force:
\begin{equation}
\vec{f}_{\rm mf}={R\over 1+R^2} \hat{\omega}_n\times (\vec{\omega}_n\times
\vec{w}_{\rm np})+
                {R^2\over 1+R^2}\vec{\omega}_n\times \vec{w}_{\rm np},
\label{mutfr1}
\end{equation}
where $\vec{\omega}_n=\nabla\times \vec{V}_n$ is the vorticity of the
neutron fluid
in a non-rotating frame
(here $\vec{V}_n$ is the neutron velocity in the non-rotating
frame)\footnote{In GAJ1,
$\vec{\omega}_n$ is erroneously defined as $\nabla\times \vec{v}_n$.
However, in their
subsequent calculations they, most likely, use the correct expression.},
$\hat{\omega}_n=\vec{\omega}_n/|\vec{\omega_n}|$ is the associated
unit vector,
$\vec{w}_{\rm np}=\vec{v}_n-\vec{v}_p$, and $R$ is the dimensionless number
measuring the strength of the drag between the neutron vortices and
the plasma.
When $R\ll 1$  (the weak-drag limit), the first term on the right-hand side
dominates. This entails that the neutron vortices mostly follow
the motion of the neutron superfluid in the direction perpendicular to
$\vec{\omega}_n$. When $R\gg 1$ (the strong-drag limit), the second
term on the right-hand side dominates. This entails that the neutron
vortices mostly
follow the plasma motion. When $R=\infty$, which is the
case on which this paper focuses, the vortices get pinned to the plasma.
In this
limit, the plasma and the neutron superfluid interact exclusively
via the Magnus force arising from the relative motion between the neutron
vortices
and neutron superfluid.

We choose the background state as follows: 1. the $z$-axis is directed along
$\Omega$; 2. the neutron vortices are aligned with $\vec{\Omega}=\Omega
\vec{e}_z$,
and are at rest in the rotating frame; 3. in the same frame,
the plasma has a background velocity $\vec{w}_0=w_0 \vec{e}_z$, which is
directed
along the vortices; 4. the mean magnetic field is
directed along the vortices, $\vec{B}=B \vec{e}_z$. We consider waves
which are
propagating along the z-axis. We are interested in
the waves for which the restoring force is the combination of
hydromagnetic stress,
the Coriolis force, and the Magnus force.
This means that the wave must be nearly incompressible, which implies
\begin{equation}
\vec{k}\cdot \vec{\delta v}_{n,p}=0.
\label{ortho}
\end{equation}
Here $\vec{k}$ is the wavevector, $\vec{\delta v}_{n,p}$ is the
neutron/proton
velocity perturbation due to the wave.
Incompressibility and assumed homogeneity of the background state imply
$\delta
\psi_{n,p}=0$.

Let us introduce the Lagrangian displacement vectors $\vec{\xi}_{n,p}$ of the
neutron and proton fluids from their
background positions,
with $\vec{\delta v}_{n,p}=D_t^{n,p}\vec{\xi}_{n,p}$.
We are looking for the solutions of the form
\begin{equation}
\vec{\xi}_{n,p}(z,t)=\left[\xi^{n,p}_{x0} \vec{e}_x+\xi^{n,p}_{y0}
\vec{e}_y\right]
e^{i(\sigma t+kz)},
\label{anzatz}
\end{equation}
where $\sigma$ is the angular frequency of the wave.

We now perturb Equations (\ref{n1}), (\ref{np}), and (\ref{mutfr1}); we set
$R=\infty$ in the latter.
To the linear order in the velocity perturbation, we have:
\begin{eqnarray}
D^n_t \vec{\delta v}_n&=& \partial^2\vec{\xi}_n/\partial t^2=-\sigma^2
\vec{\xi}_n,\nonumber\\
D^p_t \vec{\delta v}_p&=&-(\sigma+k w_0)^2\vec{\xi}_p,\label{perturb1}\\
\delta\vec{f}_{\rm mf}&=&2\vec{\Omega}\times\left(\vec{\delta v}_n-\vec{\delta
v}_p\right)-
                  \left(\nabla\times \vec{\delta
v}_n\right)\times\vec{w}_0,\nonumber\\
\nu_{\rm ee}\nabla^2\vec{\delta v}_p&=&-i\nu_{\rm ee}k^2
(\sigma+kw_0)\vec{\xi}_p,\nonumber\\
\delta\vec{f}_{\rm hm}&=&c_A^2 \partial^2\vec{\xi}_p/\partial
z^2=-c_A^2k^2\vec{\xi}_p.\nonumber
\end{eqnarray}
Here $c_A=\sqrt{B B_{\rm cr}/(4\pi \rho_p)}$ is the Alfven velocity in the
plasma. {The expression for $\delta\vec{f}_{\rm hm}$ is obtained using the magnetic induction equation.}
Substituting these into Eqs.~(\ref{n1}) and (\ref{np}), and using
$\nabla\times\vec{\xi}=i\vec{k}\times
\vec{\xi}$ together with $\vec{\delta v}_n= i\sigma \vec{\xi}_n$ and
$\vec{\delta
v}_p=i(\sigma+kw_0) \vec{\xi}_p $,
we get 2 linear vector equations for $\vec{\xi}_n$ and $\vec{\xi}_p$.
It is now convenient to proceed as follows:
%One could decompose these vectors into
%$x$ and $y$ components, and obtain 4 linear equations with 4 variables.
%One would
%then set the determinant
%of the correspondent $4\times 4$ matrix to zero and obtain the dispersion
%relation.
%We however simplify this calculation
%as follows:

Let us represent a vector $\vec{\xi}=\xi_x \vec{e}_x+\xi_y \vec{e}_y$ by a
complex
number
$\tilde{\xi}=\xi_x+i\xi_y$. Then  a vector $\vec{e}_z\times \vec{\xi}$ is
represented by $i\tilde{\xi}$.
By using this, we can immediately rewrite the 2 real vector equations as
2 complex scalar equations:
\begin{eqnarray}
\sigma\tilde{\xi}_n+2\Omega\tilde{\xi}_p&=&0,\nonumber\\
\left[\bar{\sigma}+2\Omega\left(1-{1\over x_p}\right)-\left(i\nu_{\rm
ee}+c_A^2/\bar{\sigma}\right)k^2\right]
\bar{\sigma}\tilde{\xi}_p+& &\label{linear2}\\
{2\Omega-kw_0\over x_p}\sigma \tilde{\xi}_n&=&0,\nonumber
\end{eqnarray}
where $\bar{\sigma}=\sigma+kw_0$. This pair of equation yields immediately
the
complex dispersion relation:
\begin{eqnarray}
\bar{\sigma}^2+\left[2\Omega\left(1-{1\over x_p}\right)-i\nu_{\rm
ee}k^2\right]\bar{\sigma}-& &\nonumber\\
{2\Omega(2\Omega-kw_0)
\over x_p}-c_A^2k^2&=&0.\label{maineq}
\end{eqnarray}
The dispersion relation for arbitrary $R$ is derived, for completeness, in
Appendix A.

In the next 2 sections we consider 2 applications of the relation
Eq.~(\ref{maineq}).

\section{Hydromagnetic waves in magnetars}
In this section we assume that there is no $\vec{\Omega}$-directed relative
proton-neutron flow, i.e.~we assume $w_0=0$.
We also set $\nu_{\rm ee}$ to zero, since the ratio
of the viscous to hydromagnetic stress is given by
\begin{equation}
\nu_{\rm ee}\sigma/c_A^2\ll 1.
\label{ratio}
\end{equation}
With these simplifications, the dispersion relation (\ref{maineq})  gives
 \begin{equation}
\sigma=-\Omega\left(1-{1\over x_p}\right)\pm \sqrt{\Omega^2\left(1+{1\over
x_p}\right)^2+c_A^2k^2}.
\label{sigmamag}
\end{equation}
It is important to note that in this expression $c_A$ is a function of only the proton density $\rho_p$ ($c_A^2 \equiv B B_{\rm cr}/4\pi \rho_p$).
All observed magnetars are slowly rotating, with $\Omega\sim 1$rad/s. The
observed
lowest angular frequency
for a magnetar QPO is $18$Hz, thus $\sigma \sim 113$rad/s. 
%{\bf This is in
%remarkable correspondence with the computed frequency of a plain Alfven
%fundamental wave $\nu_{\rm Alfven} = c_A^2 k^2 = 20$Hz (see
%Eq.~(\ref{nua}) and (\ref{beff}), where we use $B=10^{15}$G, $R=10$km,
%$\rho_n=10^{15}$g/${\rm cm^3}$ and $x_p=0.05$.) 
The
sum of Magnus and Coriolis forces, represented
by the terms with $\Omega$, contribute only a fraction 
 $\delta
\sigma/\sigma$ to the wave frequency, given by
\begin{equation}
\delta \sigma/\sigma\simeq {\Omega\over x_p\sigma}=0.18 \left({\Omega\over
1\hbox{rad/s}}\right)
\left({113\hbox{rad/s}\over \sigma}\right)\left({0.05\over x_p}\right).
\label{ratio1}
\end{equation}
We note that this does {\it not} depend on the assumption that
$\sigma$ represents some fundamental Alfven mode.

>From the above equation, it is clear that for hydromagnetic waves
associated with the magnetar QPO frequencies of $18$Hz and higher,
the magnus forces from neutron superfluid introduce only a small correction
to their propagation. Thus we conclude that the magnetar
oscillations (as
seen in the
giant-flare QPOs) even in the case of strong pinning, do not couple
significantly to the neutron superfluid.\footnote{We note that we have used 
the plane-wave analysis for what is likely
a non-plane-wave oscillation. This is clearly a limitation of our formalism.
However, the plane-wave analysis illustrates the physics which
is also valid for oscillatory mode of any geometry, namely that for high-frequency waves 
Alfven restoring forces are much greater than the Magnus forces.
This occurs essentially because the Magnus force is proportional to the velocity and thus 
scales as $\sigma$, while the total
restoring force scales as $\sigma^2$. Thus, we believe that our analysis is qualitatively correct in
the high-frequency regime, for non-plane-wave Alfven-type oscillations.}
Therefore, given the knowledge of the internal magnetic field, one
should use
$\rho_c\simeq x_p\rho_n$ in Eq.~(\ref{nua}) to determine the frequency of
the
lowest QPO which, according to Levin (2007), corresponds to the turning point
of the Alfven continuum in the core.
For Levin's the simplest computable magnetar model (uniform internal
magnetic field and density), with the typical magnetar parameters, $B=10^{15}$G, $R=10$km, 
$\rho=10^{15}\hbox{g}/\hbox{cm}^3$ and $x_p=0.05$,
Eq.~(\ref{nua}) gives
{$\nu_a\simeq 22$Hz, which is in qualitative agreement with the observed
$18$Hz.} 
 If
the whole
neutron superfluid would mass-load the wave, this frequency would be
reduced by a
factor of $\sim 4$. 
While suggestive, the numbers above certainly should not be taken as evidence of neutron superfluidity,
since the strength and geometry of magnetic fields inside the magnetar are highly 
uncertain.

\section{Application to precessing neutron stars}
Consider now a precessing neutron star where the neutron angular velocity
and the
crust+proton angular velocity\footnote{The crust and the core protons are
co-precessing;
this is enforced by the hydro-magnetic stresses (Levin \& D'Angelo, 2004).}
 are equal in magnitude $\Omega$  but are misaligned by an angle $\theta$.
Suppose that this angle is fixed due to the strong vortex pinning.
The relative velocity of the proton superfluid along the vortices is given by
\begin{equation}
w_0=\Omega \sin{\theta} x_2,
\label{w0}
\end{equation}
where $x_2$ is the coordinate measured along
$\vec{\Omega}_n\times\vec{\Omega}_p$.
Note that this expression agrees with Eq.~(18) in GAJ1 when one notes that
for small
$\theta$ their wobble angle $\theta_w$ equals $I_p\theta/I_n$, where $I_p$
and $I_n$
are
the moments of inertia of the proton and neutron components, respectively.

\subsection{Glaberson instability criterion}
It is convenient to express the general solution of Eq.~(\ref{maineq}) as follows:
\begin{eqnarray}
\sigma&=&-kw_0-\Omega\left(1-{1\over x_p}\right)+{i{\nu_{\rm ee}}k^2\over 2}\pm\label{mainsol}\\
      & &\sqrt{D}, \nonumber
\end{eqnarray}
where
\begin{eqnarray}
D&=&\Omega^2\left(1+{1\over x_p}\right)^2+c_A^2k^2-{2kw_0\Omega\over
x_p}-{\nu_{\rm
ee}^2k^4\over 4}-\nonumber\\
 & &i\nu_{\rm ee}k^2\Omega\left(1-{1\over x_p}\right).
\label{D}
\end{eqnarray}
This is essentially the same as Eq.~(10) of GAJ1 when $c_A=0$.

First, lets  consider the non-viscous case with $\nu_{\rm ee}=0$.
Then the minimum of $D$ is attained for $k={\Omega w_0/(x_p c_A^2)}$ and
equals
\begin{equation}
D_{\rm min}= \Omega^2\left(1+{1\over x_p}\right)^2-{\Omega^2 w_0^2\over
x_p^2 c_A^2}.
\end{equation}
Thus the {Glaberson} instability appears only for
\begin{equation}
w_0>c_A(1+x_p)\simeq c_A.
\label{criterion}
\end{equation}

We now allow for the viscous term in Eqs.~(\ref{mainsol}) and (\ref{D}).
We find that a weak, viscosity-driven  instability  appears at a smaller
velocity\\
\begin{equation}
w_0 > 2\sqrt{x_p}c_A = \sqrt{\frac{BB_{\rm cr}}{\pi \rho_n}},
\label{w0crit1}
\end{equation}
for the wave-vector range
\begin{equation}
k_-<k<k_+
\label{krange}
\end{equation}
where
\begin{equation}
k_{\pm}={\Omega\over x_p c_A^2}\left[w_0\pm\sqrt{w_0^2-4c_A^2x_p}\right];
\label{kpm}
\end{equation}
see Appendix B for the mathematical details.\\

Equation (\ref{w0crit1}) expresses the lowest relative proton-neutron velocity which
is required for the initiation of the Glaberson instability. Whether this velocity is
acheived depends on the misalignment angle $\theta$ between the proton and neutron anglular velocity
vectors. In the next subsection, we derive a simple but rigorous upper bound on $\theta$.

\subsection{The maximum misalignment angle and relative flow velocity for fast precession.}
The misalignment angle $\theta$ can be constrained, by requiring that 
the precessional torque $\tau_p$ of the proton component not exceed the maximum torque $\tau_m$ that the 
vortices can exert on the fluxtubes through magnetic stresses. The precessional torque is given by\footnote{Since 
the neutrons are pinned to the protons, the torque acting on the neutrons is given by the 
cross product of the instantaneous angular velocity of the protons and the neutrons angular momentum, 
and is therefore independent of $x_p$. In our derivation we assume 
that the angular velocities of the protons and the neutrons have the same magnitude $\Omega$.}
\begin{equation}
\tau_p = L_n \Omega_{p} \sin{\theta}
\label{torque1}
\end{equation}
Here $L_n=I\Omega$ is the proton angular momentum, $I$ is the total stellar moment of inertia and $\Omega_{p}=\Omega$. We find that
for a typical neutron star with the mass of $M=1.4 M_{\odot}$ and radius of
$R=10^6$ cm, the precessional torque is given by 
\begin{equation}
\tau_p\simeq 4\cdot 10^{46} \sin\theta (P/1\hbox{s})^{-2} \hbox{g}\hbox{cm}^2\hbox{s}^{-2}.
\label{torque11}
\end{equation}
Here $P$ is the neutron-star rotational period.
 
The maximal physically-admissible torque $\tau_m$ can be computed by
assuming that the vortices have maximal contact with the fluxtubes,
i.e.~that the vortex is pushed/pulled with the stress of $B_{\rm cr}^2/(8\pi)$ accross its whole side surface.
The maximal torque exerted on a single vortex
is given by
\begin{equation}
\tau_v={B_{\rm cr}^2\over 8\pi} l^2d,
\label{tauv}
\end{equation}
where   $d \sim 2 \cdotp 10^{-12}$ cm is the vortex diameter, and $l$ is the vortex half-length.
The total number of vortices is given by 
\begin{equation}
N=\pi R^2 n_v \sim 3 \cdotp 10^{16} (R/10^6\hbox{cm})^2 (P/1\hbox{s})^{-1},
\label{vortices}
\end{equation}
where $n_v$ is the per-area vortex density (Shapiro and Teukolsky 1983, Link 2003).
For a spherical star with a dense grid of the linear vortices, the average value of $l^2$ is $R^2/2$.
Thus, by combining Eqs.~(\ref{tauv}) and (\ref{vortices}), we arrive to the following expression:
\begin{equation}
\tau_m={B_{\rm cr}^2\over 16}{R^4d}n_v\simeq 1.3\cdot 10^{45}(P/1\hbox{s})^{-1} \hbox{g}\hbox{cm}^2\hbox{s}^{-2}.
\label{taum}
\end{equation}

>From Eqs.~(\ref{taum}) and (\ref{torque11}), we see that our requirement  $\tau_m > \tau_p$  implies that 
\begin{equation}
\theta < 2^o (P/1\hbox{s}),
\label{thetaup}
\end{equation} 
and therefore 
\begin{equation}
w_0\sim \Omega\theta R < 2 \cdotp 10^5 \hbox{cm/s}.
\end{equation}
This upper limit on $w_0$ is spin-independent.  

So is this velocity sufficient to drive the Glaberson instability?
Equation (\ref{w0crit1}) tells us that for $x_p=0.05$, $B=10^{12} \hbox{G}$ and $\rho_n=10^{15}\hbox{g}/\hbox{cm}^3$,
the critical relative velocity is $w_0\sim 6\cdot 10^5$cm/s. 
Thus we conclude that in the presence of strong vortex pinning and magnetic fields $B>10^{11} \hbox{G}$ 
the misalignment between the proton- and neutron angular velocities is unlikely to become large enough 
to provide wind velocities $w_0$ that are sufficient to drive the Glaberson instability.

\section{Outlook}
The calculations of this paper have 2 main astrophysical implications.
First, we have shown that the Alfven waves which are associated with magnetar
QPOs are not significantly mass-loaded by a neutron superfluid, even if the
superfluid vortices are strongly pinned to the proton-electron plasma.
For $B=10^{15}$G and the simplest B-field geometry, the expected frequency
of lowest
magnetar QPO is in remarkable agreement with observations, {\it if} only the
protons, i.e.~about $0.05$ of the core mass, are mass-loading the Alfven
waves.
Strong vortex pinning will, however, have a strong effect on the Alfven waves
in rapidly spinning and
relatively non-magnetic neutron stars, i.e.~those ones in the Low-Mass
X-ray Binaries.
In these stars the Alfven waves may play an important role in damping of the
r-mode instability, as discussed by Mendell (2001) and Kinney and Mendell
(2003) for
the cases of non-superfluid and superfluid core, respectively. Kinney and
Mendell
(2003)
had not considered the vortex pinning (see also Mendell 1991); however from
Eq.~(\ref{ratio1}) and from the
fact that $\sigma\sim\Omega$ for an R-mode, it is clear that the strong
vortex pinning
would have a huge (of order $1/x_p$) effect on the Alfven waves with the
R-mode
frequency.

Second, we have shown that the hydromagnetic stresses generally suppress the
Glaberson instability
in the proton-neutron superfluid mixture, in the case of strongly pinned
vortices\footnote{We have not considered here  the PMGO case when some normal neutron
component is present and
is driving the instability.}. 
We have also provided a robust upper bound Eq.~({\ref{thetaup}) on the angle between proton and neutron 
angular velocities in the fast-precessing neutron stars. Even for slowly-spinning magnetars, the misalignment
angle cannot exceed $20$ degrees, which implies a wobble angle no greater than 1 degree. 
Thus a detection of neutron-star fast precession is difficult, if not impossible, observationally.
An inspection of the XMM timing data on known anomalous x-ray pulsars produces no
statistically-significant periodic signal which could be interpreted as fast precession
[van Kerkwijk 2008, private communications].

\section{Acknowledgements}
We thank Andrew Melatos and Kostas Glampedakis for useful discussions, the anonymous referee
for valuable suggestions. We thank Maarten van Kerkwijk for a search of fast magnetar precession
in the XMM data.

\section*{References}
\begin{footnotesize} \noindent
Alpar, M.~A., Langer, S.~A., \& Sauls, J.~A.~1984, ApJ, 282, 533\\
Andersson, N., Comer, G.L., Glampedakis, K., 2005, Nucl. Phys. A 763, 212\\
Barat, C., et al., 1983, A\&A, 126, 400\\
Bekarevich, I.~L., \& Khalatnikov, I.~M., 1961, Sov.~Phys.~JETP, 13, 643\\
Cutler, C, \& Lindblom, L., 1987, ApJ, 314, 234\\
Donnelly, R.~J., 1991, ``Quantized Vortices in Helium'', Vol. II, Cambridge
University Press, Cambridge, England\\
Easson, I, \& Pethick, C.~J., 1977, Phys.~Rev.~D., 16, 265\\
Flowers, E., \& Itoh, N., 1979, ApJ, 230, 847\\
Glaberson, W.~I., Johnson, W.~W., \& Ostermeier, R.~M.~1974,
Phys.~Rev.~Lett.,
33, 1197\\
Glampedakis, K., Samuelsson, L., \& Andersson, N.~2006, MNRAS, 371, L74\\
Glampedakis, K., Andersson, N., \& Jones, D.~I.~2007, Phys.~Rev.~Lett.,
accepted
(arxiv0708.2693, GAJ1 in the paper)\\
Glampedakis, K., Andersson, N., \& Jones, D.~I.~2008, arXiv:0801.4638
(GAJ2)\\
Hall, H.~E., \& Vinen, W.~F.~1956, Proc.~R.~Soc.~Lond., A238, 204\\
Israel, G., et al.~2005, ApJ, 628, L53\\
Kinney, J.~B., \& Mendell, G.~2003, Phys.~Rev.~D., 67b4032\\
Lee, U., 2007, accepted to MNRAS (arXiv:0710.4986)\\
Levin, Y., 2006, MNRAS Lett.,368, 35\\
Levin, Y., 2007, MNRAS, 357, 159 (L07 in the paper)\\
Levin, Y., \& D'Angelo, C.~2004, ApJ, 6163, 1157\\
Link, B.~2003, Phys.~Rev.~Lett., 91, 101101\\
Link, B.~2007, Ap\&SS, 308, 435\\
Lyutikov, M., 2003,  MNRAS, 346, 540\\
Mendell, G.~1991, ApJ, 380, 515\\
Mendell, G.~2001, Phys.~Rev.~D., 64d4009\\
Melatos, A., \& Peralta, C.~2007, ApJ Letters, 662, 99\\
Pedlosky, J., 1979, ``Geophysical fluid dynamics'' (New York, Springer)\\
Peralta, C., Melatos, A., Giacobello, M., \& Ooi, A.~2005, ApJ, 635, 1224
(PMGO5)\\
Peralta, C., Melatos, A., Giacobello, M., \& Ooi, A.~2006, ApJ, 651, 1079
(PMGO6)\\
Ruderman, M., Zhu, T., \& Chen, K.~1998, ApJ, 492, 267\\
Sedrakian, A., Wasserman, I., \& Cordes, J.M., 1999,  ApJ, 524, 341\\
Sidery, T., Andersson, N., \& Gomer, G.~L.~2007,
submitted to MNRAS, arxiv0706.0672 (SAG in the paper)\\
Shaham, J.~1977, ApJ, 214, 251\\
Shapiro, S.L., \& Teukolsky, S.A., 1983, ``Black holes, white dwarfs, and
neutron
stars:
the physics of compact objects'' (New York, Wiley-Interscience)\\
Sotani, H., Kokkotas, K., \& Stergioulas, N., 2007, accepted to MNRAS
(arXiv:0710.1113)\\
Strohmayer, T.E., \& Watts, A.L., 2005, ApJ, 632, L111\\
Strohmayer, T.E., \& Watts, A.L., 2006, ApJ, 653, 594\\
Thompson, C., \& Duncan, R.C., 1995, MNRAS, 275, 255\\
\end{footnotesize}

\appendix
\section{Dispersion relation for arbitrary drag}

In this Appendix we perform a plane-wave analysis of the two-fluid
dynamical equations
defined in eq. (3) and (4), and derive a dispersion relation for arbitrary
drag strength $R$.
Using the plane wave solutions from equation (7), we can perturb equations
(3) and (4). Retaining linear terms we get:
\begin{equation}
D_t^n \delta v_n = - \sigma^2 \vec{\xi}_n
\end{equation}
\begin{equation}
D_t^p \delta v_p = -\bar{\sigma}^2 \vec{\xi}_p
\end{equation}
\begin{equation}
2 \delta \vec{v}_n \times \vec{\Omega} = 2 i \Omega \sigma \vec{\xi}_n
\times \vec{e}_z
\end{equation}
\begin{equation}
2 \delta \vec{v}_p \times \vec{\Omega} = 2 i \Omega \bar{\sigma}
\vec{\xi}_p \times \vec{e}_z
\end{equation}
\begin{eqnarray}
\delta \vec{f}_{\rm mf} = \frac{R}{1+R^2} \left[ 2 i \Omega \vec{e}_z
\times \left ( \vec{e}_z \times \left( \sigma \vec{\xi}_n - \bar{\sigma}
\vec{\xi}_p \right) \right) + k w_0 \sigma \vec{e}_z \times \left( \left(
\vec{e}_z \times \vec{\xi}_n \right) \times \vec{e}_z \right) \right]+ \\
\frac{R^2}{1+R^2} \left[2 i \Omega \vec{e}_z \times \left( \sigma
\vec{\xi}_n -
\bar{\sigma} \vec{\xi}_p \right) - w_0 k \sigma \vec{e}_z \times \left(
\vec{e}_z \times \vec{\xi}_n \right) \right] \nonumber
\end{eqnarray}
\begin{equation}
\nu_{\rm ee} \nabla^2 \delta \vec{v}_p = - \nu_{\rm ee} k^2 \bar{\sigma}
\vec{\xi}_p
\end{equation}
\begin{equation}
\delta \vec{f}_{\rm hm} = - c_A^2 k^2 \vec{\xi}_p
\end{equation}

Here $\bar{\sigma} \equiv \sigma + w_0 k$ and $c_A = \sqrt{BB_{\rm
cr}/4\pi \rho_c}$ is the Alfven velocity in the plasma. We can simplify
these expressions by using the same trick as in Section 2: Let us
represent the
vector $\vec{\xi} = \xi_x \vec{e}_x + \xi_y \vec{e}_y$ by a complex number
$\tilde{\xi} = \xi_x + i \xi_y$. The cross product $\vec{e}_z \times
\vec{\xi}$ corresponding to a simple rotation in the $xy$-plane, can then
be represented by $i\tilde{\xi}$. Using this, we convert the two real
vector equations (3) and (4) into two complex scalar equations:

\begin{equation}
-2\Omega C \bar{\sigma} \tilde{\xi}_p = \left[ \bar{\sigma} + \left( C-1
\right) \left( w_0 k - 2 \Omega \right) \right] \sigma \tilde{\xi}_n \\
\end{equation}
\begin{equation}
\left[ \bar{\sigma} + \left( 2 \Omega \left( 1 - \frac{C}{x_p}\right) - i
\nu_{\rm ee} k^2 \right) - \frac{c_A^2}{\bar{\sigma}} k^2 \right]
\bar{\sigma}\tilde{\xi}_p = \frac{C}{x_p} \left( w_0 k - 2\Omega \right)
\sigma \tilde{\xi}_n
\end{equation}

where $C \equiv \frac{R\left( i + R \right)}{1 + R^2}$. This pair of
equations yields the following complex dispersion relation:

\begin{eqnarray}
\bar{\sigma}^3 + \bar{\sigma}^2 \left[ \left( C-1 \right) \left( w_0 k -
2\Omega \right) + 2\Omega \left( 1-\frac{C}{x_p} \right) - i \nu_{\rm ee}
k^2 \right] +\label{maineq2}\\
 \bar{\sigma} \left[ \left( w_0 k - 2\Omega \right) \left( \left(
C-1\right) \left( 2\Omega - i \nu_{\rm ee} k^2 \right) + \frac{2\Omega
C}{x_p} \right) - c_A^2 k^2 \right] - c_A^2 k^2 \left( C-1 \right) \left(
w_0 k - 2 \Omega \right) = 0 \nonumber
\end{eqnarray}

In the strong drag limit $R \rightarrow \infty$ ($C=1$) this cubic
equation simplifies significantly:

\begin{equation}
\bar{\sigma}^2 + \bar{\sigma} \left[ 2\Omega \left( 1-\frac{1}{x_p}
\right)-i\nu_{\rm ee} k^2 \right] - \left[ \frac{2\Omega \left( 2\Omega -
w_0 k \right)}{x_p} + c_A^2 k^2 \right] = 0,
\end{equation}
which is Eq.~(\ref{maineq}) in the text.

\section{Instability criterion for non-zero viscosity}
In this Appendix we derive a criterion for instability in the case of
non-negligible viscosity, i.e.~Eq.~(\ref{w0crit1}). We rewrite
Eq.~(\ref{mainsol}) as follows:

\begin{equation}
\sigma = A + iB \pm \sqrt{C+iD}\label{B1}
\end{equation}

where
\begin{equation}
A=-w_0 k - \Omega \left( 1 - \frac{1}{x_p} \right) \nonumber
\end{equation}
\begin{equation}
B= \frac{ \nu_{\rm ee} k^2}{2}\nonumber
\end{equation}
\begin{equation}
C=\Omega^2 \left( 1+\frac{1}{x_p} \right)^2 + c_A^2 k^2 - \frac{2 w_0 k
\Omega}{x_p} - B^2\label{B2}
\end{equation}
\begin{equation}
D= - 2\Omega \left( 1-\frac{1}{x_p}\right) B \nonumber
\end{equation}
The state of marginal stability is given by
\begin{equation}
Im \left( \sigma \right) = B \pm Im
\left( \sqrt{C+iD} \right) = 0.
\label{marginal}
\end{equation}
In order to find a convenient expression
for $Im \left( \sqrt{C+iD} \right)$ we write $ \left( \sqrt{C+iD} \right)$
in polar form. The imaginary part is then given by \\

\begin{equation}
Im \left( \sqrt{C+iD} \right) = \left( C^2 + D^2 \right)^{1 \over 4}
\sin{\left( \frac{1}{2} \arccos{\left( \frac{C}{\sqrt{C^2 + D^2}} \right)
} \right)}
\label{B3}
\end{equation}

Combining this with Eq.~{\ref{marginal}), we find

\begin{equation}
B^2 = \frac{1}{2} \sqrt{C^2 + D^2 } - \frac{1}{2} C
\end{equation}

Taking the square of this expression, and using Eqs (\ref{B2}) we arrive at

\begin{equation}
\frac{2\Omega \left( 2\Omega - w_0 k \right)}{x_p} + c_A^2 k^2 = 0
\end{equation}

And therefore,

\begin{equation}
k_{\pm} = \frac{\Omega}{x_p c_A^2} \left[ w_0 \pm \sqrt{w_0^2 - 4 c_A^2 x_p}
\right]
\end{equation}

The  unstable waves have wave-vectors in the interval $k_-<k<k_+$,
provided that $k_-$ and $k_+$ are
real. Thus the criterion for the instability is

\begin{equation}
w_{0} > 2 c_A \sqrt{x_p},
\end{equation}

this is Eq.~(\ref{w0crit1})
of the paper.  We now make an estimate of the growth rate in this instability window.
For a realistic neutron star, we take
$\nu_{\rm ee} \approx 10^9 \hbox{cm}^2\hbox{s}^{-1}$ at $T \approx
10^7\hbox{K}$
(Flowers \& Itoh 1979, Cutler \& Lindblom 1987, Andersson, Comer \&
Glampedakis, 2005),
$\Omega \approx 2\pi\hbox{rad/s}$, $c_A \approx 10^6\hbox{cm/s}$ and
therefore for $w_0\sim c_A$,
the wave-vector of an unstable wave is $k\sim 10^{-4}\hbox{cm}^{-1} $. We
now note that $\nu_{\rm ee} k^2 /2 << \Omega /x_p$.
Therefore the terms
$B$ from Eq.~(\ref{B1}) and $B^2$ from Eq.~(\ref{B2})
 have a negligible contribution to $Im \left( \sigma \right)$ and can
be ignored. \\

First the case where $C < 0$. We note
that $Im \left( \sigma
\right)$ is completely dominated by $C$ so that we
arrive at the criterion of Eq.~(\ref{criterion}) again. With a growth rate of\\

\begin{equation}
-Im\left( \sigma \right) = \sqrt{C}
\label{eq}
\end{equation}

Next consider $C > 0$, i.e. $w_0 < \left( 1 + x_p
\right) c_A$. By means of a simple analysis in the complex plane one can
show that \\

\begin{equation}
\frac{\sqrt{D}}{2}
< - Im \left( \sigma \right) < \sqrt{\frac{D}{2}}
\label{eq22}
\end{equation}
For $w_0 > 2 \sqrt{x_p}c_A$, there is a range of $k$ where
the instability occurs; see Eq.~(\ref{krange}). Substituting $k=k_+$ into
Eq.~(\ref{eq22}) for the
maximum $k$ in this range, we find the growth rate of the instability
\begin{equation}
- Im \left( \sigma \right) \approx \sqrt{\frac{\nu_{\rm ee}\Omega^3}{2 x_p^3 c_A^4}}\left[ w_0 + \sqrt{w_0^2 - 4 c_A^2 x_p} \right]
\end{equation}

We note that because the viscosity is small, this growth rate is much smaller than the one in Eq.~(\ref{eq}).
\end{document}